\documentclass[onecollarge,natbib]{svjour2}
\bibpunct{[}{]}{;}{n}{}{,} 
\smartqed  
\usepackage{graphicx}
\usepackage{color} 
\usepackage{bm}       
\usepackage{amsmath}  
\usepackage{amssymb}
\definecolor{purple}{rgb}{0.5,0,0.5}
\definecolor{blue}{rgb}{0.0,0,0.9}
\usepackage[colorlinks=true, pdfstartview=FitV, citecolor= purple, linkcolor = blue,urlcolor=blue]{hyperref}

\newcommand{\feynslash}[1]{\slash\hspace*{-1.8mm} #1}

\journalname{Few-Body Systems}
\begin{document}

\title{Excited hadrons and the analytical structure of bound-state interaction kernels }


\author{\mbox{Bruno El-Bennich \and Gast\~ao Krein \and Eduardo Rojas  \and Fernando E.~Serna}}


\institute{Bruno El-Bennich   \at
                Laborat\'orio de F\'isica Te\'orica e Computacional, Universidade Cruzeiro do Sul, Rua Galv\~ao Bueno 868, 01506-000 S\~ao Paulo, SP, Brazil.
                \email{bruno.bennich@cruzeirodosul.edu.br}     
                \and
                Gast\~ao Krein and Fernando Serna  \at Instituto de F\'isica Te\'orica, Universidade Estadual Paulista, Rua Dr. Bento Teobaldo Ferraz, 271 -- Bloco II, 01140-070 S\~ao Paulo, 
                SP, Brazil. \email{gkrein@ift.unesp.br ; fernandoserna@ift.unesp.br}
                \and 
                Eduardo Rojas \at Instituto de F\'isica, Universidad de Antioquia, Calle 70, no.~52-21, Medell\'in, Colombia. \email{\mbox{rojas@gfif.udea.edu.co}}}

\date{Received: date / Accepted: date}

\maketitle

\begin{abstract}
We highlight Hermiticity issues in bound-state equations  whose kernels are subject to a highly asymmetric mass and momentum distribution and whose eigenvalue spectrum 
becomes complex for radially excited states. We trace back the presence of imaginary parts in the eigenvalues and wave functions to truncation artifacts
and suggest how they can be eliminated in the case of charmed mesons. The solutions of the gap equation in the complex plane, which play a crucial role in
the analytic structure of the Bethe-Salpeter kernel, are discussed for several interaction models and qualitatively and quantitatively compared to analytic continuations 
by means of complex-conjugate pole models fitted to real solutions.

\end{abstract}

\section{Introduction}

A time-honored  tool to study and understand the structure of matter and its constituents, be it solids, molecules, atoms or nuclei, is spectroscopy. Indeed, if we revisit the 
history of atoms, it becomes clear that spectroscopy has been an invaluable and very efficient tool to reveal   their {\em quantified\/} nature via spectral lines and 
to understand their constituent content. Eventually, the discoveries made thanks to spectroscopy led  to the foundation of quantum theory and its refinement to the development 
of Quantum Electrodynamics with its glorious explanation, amongst other phenomena, of the Lamb shift~\cite{Tomonaga:1946zz,Schwinger:1948iu,Feynman:1949hz,Dyson:1949bp}. 

A like-minded approach has been pursued in particle and hadron physics  since the early experiments at the Stanford Linear Accelerator which ultimately led to the
quark picture of hadrons. The quark model was eventually promoted to a non-Abelian gauge theory we nowadays believe to describe the strong interactions between 
the quarks, namely Quantum Chromodynamics (QCD)~\cite{Fritzsch:1973pi}. 
In recent years this program has been pursued and strongly extended with meson electroproduction experiments~\cite{Aznauryan:2012ba} at Jefferson Lab where
its Continuous Electron Beam Accelerator Facility (CEBAF) recently delivered the first batch of 12~GeV electrons to Hall D. In particular, the extensive study of mesons and baryon 
ground states, the mass differences and level ordering with their radially excited states and parity partners as well as with exotics shed important light on the constituent 
structure of the hadrons. In some cases it will help to distinguish between very different or controversial pictures of a hadron's composition, for instance the competing 
descriptions of scalar mesons and the $X(3872)$ as either molecular bound states of lighter mesons and tetraquarks; we note that quantum field theory does not prevent 
these hadrons to be a superposition of many-quark states {\em and\/} flavored meson loops.

At a fundamental level, we are not merely interested in spectroscopy as a means to inform us about the mass spectrum and level orderings of hadrons. Confinement
and asymptotic freedom are the prime paradigms of QCD and we expect the experimental data to teach us something about the interaction that holds the quarks together 
in bound states and in particular at larger distances or lower momenta. This is because if confinement is related to the analytic properties of QCD's Schwinger functions, 
we should gain more insights into its mechanism by mapping out the infrared behavior of  the  theory's  coupling constant. This task cannot be completed by perturbative 
analyses, yet an adequate nonperturbative continuum approach is provided by Dyson-Schwinger equations (DSE)~\cite{Dyson:1949ha,Schwinger:1951ex,Roberts:1994dr,
Alkofer:2000wg}. More precisely, one compares the effect of DSE predictions for the quark's mass functions embedded in bound-state calculations with the hadronic mass 
spectrum as well as form factors, wherefrom one obtains valuable information on the long-range behavior of the strong interaction~\cite{Bashir:2012fs,Rojas:2013tza,
El-Bennich:2013yna,Rojas:2014tya,El-Bennich:2015kja,Aguilar:2010cn,Cloet:2013jya}. 

We remind that excited hadron properties are considerably more sensitive to this long-range behavior than those of ground states~\cite{Holl:2004fr,Holl:2005vu,Hilger:2014nma,
Hilger:2015ora,Qin:2011xq,Rojas:2014aka}.  Therefore, excited mesons and nucleons~\cite{Aznauryan:2012ba} provide an extremely valuable source of improving our understanding 
of QCD in the strong coupling regime, additional to the insight gained from light mesons \cite{Maris:1997hd,Maris:1997tm,Maris:1999nt,ElBennich:2008qa,Chang:2009ae,daSilva:2012gf,
ElBennich:2012ij,deMelo:2013zza,Eichmann:2008ae}, nucleons~\cite{Oettel:2000jj,Cloet:2008re,Eichmann:2009qa,Eichmann:2011ec,Eichmann:2011aa,Segovia:2014aza,Segovia:2016end,
Sanchis-Alepuz:2014wea} and heavy-light mesons characterized by a disparate range of energy scales~\cite{ElBennich:2009vx,ElBennich:2010ha,ElBennich:2011py,ElBennich:2012tp,
ElBennich:2009da,Gomez-Rocha:2015qga,Serna:2015,Krein:2014zaa}.

In here, we focus on nontrivial issues which arise out of artifacts of a given truncation scheme of the gap and bound-state equations. The latter is given by the Bethe-Salpeter 
equation (BSE). In particular the simplest truncation that satisfies the axialvector Ward-Green-Takahashi identity (WGTI), the rainbow-ladder truncation, produces a 
non-Hermitian interaction kernel once one departs from (relatively) symmetric mass constellations, such as pions, kaons, $\eta$, $\rho$ and quarkonia. We describe in 
Section~\ref{sec2} under which conditions this non-Hermiticity manifests itself and argue that kernels beyond the rainbow-ladder truncation are indispensable once the physical 
bound states are not protected by the WGTI, as it is the case for Goldstone bosons, or the gauge-fermion vertex in the simplified  infinite-mass limit becomes invalid. 
Calculations of BSE beyond the leading symmetry-preserving truncation are underway and as a digression we discuss in Section~\ref{sec3} the complex-conjugate pole approach 
that has been employed to represent the quark propagators in the complex plane~\cite{Bhagwat:2002tx} for approaches in and beyond the rainbow-ladder truncation.

\section{Interaction kernel and hermiticity issues \label{sec2}}

As we work in continuum QCD, the interactions of a quark-antiquark pair are described by a BSE which is treated as an eigenvalue problem. As an example, consider the
relativistic bound-state equation of a pseudoscalar $J^P= 0^-$  meson with relative $\bar qq$ momentum $p$ and total momentum  $P$ omitting flavor and Dirac indices 
for simplicity: 
\begin{equation}
\label{BSE} 
  \Gamma_{0^-} (p,P) = \int \!\frac{d^4k}{(2\pi)^4} \  \mathcal{K} (p,k,P)\left  [ S (k + \eta_+P)\, \Gamma_{0^-}  (k,P)\, S (k-\eta_- P) \right ]  \ .
\end{equation}
The dressed quark propagators $S (k\pm \eta_\pm P)$ are solutions of the DSE for a given flavor with  $\eta_+ +\eta_- =1$, where $\eta_\pm$ are arbitrary partition parameters
since numerical  results are independent of the momentum distribution in a Poincar\'e invariant calculation. In rainbow-ladder truncation, the interaction kernel is given by,
\begin{equation}
\label{BSEkernel} 
   \mathcal{K} (p,k,P) =  - \frac{Z_2^2\, \mathcal{G} (q^2 )}{q^2} \, \frac{\lambda^a}{2}  \gamma_\mu  T_{\mu\nu}(q) \,
    \frac{\lambda^a}{2}  \gamma_\nu  \  ,
\end{equation}
which introduces the transverse projection operator, $T_{\mu\nu} (q) :=  g_{\mu\nu} -  q_\mu q_\nu/q^2 $, with $q=p-k$, and where $Z_2$ is the wave-function renormalization
constant and $\lambda^a$ are the SU(3) color matrices in the fundamental representation. 

A series of ans\"atze for the effective interaction $\mathcal{G} (q^2 )$ has been studied~\cite{Maris:1997hd,Maris:1997tm,Maris:1999nt,Qin:2011dd}, which serves to emulate 
the combined effect of the gluon and quark-gluon vertex dressing functions and reflects the historic evolution of the understanding of the infrared limit of the strong interaction.
There have been various efforts to extend these models by including other tensor components of the vertex beyond the leading truncation~\cite{Matevosyan:2006bk,Chang:2009zb,
Chang:2011ei,Fischer:2009jm,Qin:2013mta,Aguilar:2014lha,Binosi:2014aea,Binosi:2016rxz,Sanchis-Alepuz:2015qra}. An important feature of Eq.~(\ref{BSEkernel}) is that it satisfies 
the axialvector WGTI~\cite{Bender:1996bb} which warrants a massless pion in the chiral limit. The set of Eqs.~(\ref{BSE}) and (\ref{BSEkernel}) defines an eigenvalue problem with 
physical solutions at the mass-shell points,  $P^2 = -M_i^2$, where $M_i^2$ is the bound-state mass of the ground and excited states; see  discussions in 
Refs.~\cite{El-Bennich:2015kja,Krassnigg:2003wy}.

Using a Euclidian nonorthogonal basis with respect to the Dirac trace, the Poincar\'e invariant solutions of Eq.~(\ref{BSE}) are generally given by:
\begin{equation}    
  \Gamma_{0^-} \, (p, P)  =  \gamma_5\Big [\, i\, \mathbb{I}_D  E_{0^-}  (p,P) +  \feynslash P F_{0^-}  (p,P)
                                        +\ \feynslash p (p\cdot P)\, G_{0^-}  (p,P) + \sigma_{\mu\nu}\, p_\mu P_\nu\,   H_{0^-}  (p,  P) \, \Big ]  \ , 
\label{diracbase}                            
\end{equation}
The scalar functions $\mathcal{F}_{0^-} ^\alpha (p,P) =$  $\big \{ E_{0^-}  (p,P), F_{0^-}  (p,P), G_{0^-}  (p,P), H_{0^-}  (p,P)\big \}$ are the Lorentz invariant 
components of the Bethe-Salpeter amplitude and it can be shown~\cite{El-Bennich:2015kja} that the BSE rewritten in component form,
\begin{equation}
  \lambda(P^2)\, \mathcal{F}_{0^-} ^\alpha (p,P) =\int \!\frac{d^4k}{(2\pi)^4} \ \mathcal{K}^{\alpha\beta} (p,k,P)\, \mathcal{F}_{0^-} ^\beta (k,P) \ ,
 \label{eigenvalue}
\end{equation}
is an eigenvalue equation as indicated on the left-hand side of Eq.~(\ref{eigenvalue}) by the  eigenvalue function $\lambda(P^2)$. The latter has a solution for every value of $P^2$.
The tensor $\mathcal{K}^{\alpha\beta} (p,k,P)$ is a projection of Eq.~(\ref{BSE}) with the interaction kernel in Eq.~(\ref{BSEkernel})~\cite{El-Bennich:2015kja,Rojas:2014aka}.
The scalar functions $\mathcal{F}_{0^ -}^\alpha (p,P)$ in Eq.~(\ref{eigenvalue}) are expanded in Chebyshev polynomials of the second kind, $U_m(z_k)$ and $U_m(z_p)$, 
which are functions of the angles, $z_k = P\!\cdot k/(\surd{P^2}\surd{k^2})$ and $z_p = P\!\cdot p/(\surd{P^2}\surd{p^2})$, for example:
\begin{equation}
  \mathcal{F}_{0^-}^\alpha (p,P) =\sum_{m=0}^\infty \mathcal{F}_{0^-}^{\alpha m} (p,P)\, U_m(z_p) \ .
  \label{chebyshev}
\end{equation}
This expansion can of course also be applied to mesons with quantum numbers other than $0^{-+}$.

One of the postulates of quantum mechanics states that the eigenvalues that describe a physical observable must be real, hence the necessity of Hermitian operators. Equivalently, in case of 
the BSE,  the eigenvalue spectrum should be positive definite~\cite{Ahlig:1998qf} owing  to the Hermiticity requirement of physical operators. The eigenvalues in Eq.~(\ref{eigenvalue})
describe a mass spectrum with physical solutions for $\lambda_0(M_0^2) = \lambda_1(M_1^2) = \lambda_2(M_2^2) =  \cdots =  \lambda_i(M_i^2) =1$, where $M_0^2$ denotes the ground 
state and  $M_i^2$, $i=1,2, ...$ are radial excitations with $M_i < M_{i+1}$, and are ordered as $\lambda_0(M_i^2) >  \lambda_1(M_i^2) > \cdots > \lambda_i(M_i^2)$. As an example, the 
individual eigenvalue trajectories, $\lambda_0(M^2)$ and $\lambda_1(M^2)$, for the nucleon and its first excited state are visualized in Figure~1 of Ref.~\cite{El-Bennich:2015kja}. 

Nonetheless, within the framework of the rainbow-ladder truncation using an infrared-massive and finite interaction in agreement with numerical results from lattice QCD and DSE  for the 
gluon dressing function, we observe that the eigenvalues of heavy-light systems become complex with increasing mass difference of the antiquark-quark pairs~\cite{Rojas:2014aka}. 
More precisely, we find a  complex-conjugate pair of eigenvalues for the first radially excited states of $D$  and $D_s$ mesons when we include higher Chebyshev polynomials, $U_{m>1} (z_p)$. 
The ground-state eigenvalues of the $D$ mesons remain real. This is independent of the amount  of Chebyshev moments employed and one finds that for  $U_m (z_p), m \geq 4$,  the 
numerical solution always converges to the same eigenvalue. The appearance of imaginary parts in higher eigenvalues is not limited to mesons --- we also find them in case 
of the Faddeev kernel that describes the three-quark correlation function with the quantum numbers of the nucleon. Its first radial excitation, hence the second eigenvalue the Faddeev 
kernel produces, was shown to be consistent with the Roper resonance~\cite{Segovia:2015hra}. However, the third eigenvalue which corresponds to the second excited state also 
consists of a complex-conjugate pair. 

The above findings indicate that the rainbow-ladder kernels, which describe the repeated antiquark-quark interactions, are not Hermitian when one of the quarks becomes much heavier. 
Whereas the eigenvalues and Chebyshev moments of the kaon and its excited states are all real, we observe a threshold midway between the strange and charm masses
where the eigenvalues acquire an imaginary component. Moreover, in the case of $D$ and $D_s$ mesons we observe that {\em odd\/} Chebyshev moments contribute and they 
acquire an imaginary part in the ground and excited states. The odd contributions are not surprising since $K$ and $D$ mesons are not eigenstates of the  charge-conjugation operator 
and thus $\bar \Gamma (k,P) = \lambda_c \Gamma (k,P)$ does not imply $\lambda_c =\pm 1$ for the charge parity, whereas for  equal-mass pseudoscalar mesons with $J^{PC} = 0^{-+}$, 
the  constraint that  the  Dirac base  satisfies $\lambda_c = + 1$ imposes that $\mathcal{F}_{0^-}^\alpha (p,P)$ be {\em even\/} in the angular variable $z_p$. Yet, the angular 
dependence in the odd moments is the source of their complexity when the mass difference increases --- the zeroth Chebyshev moment of the $D$ mesons is angle independent
and remains real. This does not occur for the kaon where flavor-symmetry and charge-parity breaking are still negligible.

Thus, the appearance of complex solutions of the BSE for heavy-light mesons is intimately related with the asymmetric momentum distribution within the bound state at a given
truncation. Consider initially a pseudoscalar meson in the isovector channel with equal masses in the rainbow-ladder truncation. A consistent next-order truncation of the DSE and
BSE implies vertex-corrections and crossed-box contributions in the BSE kernel~\cite{Bender:1996bb}. A necessary consequence of the Goldstone theorem is that  these contributions
cancel, which is the reason for the successes of the rainbow-ladder truncation in many applications in hadron physics. Now, consider again a pseudoscalar isovector meson but with two 
very different quark masses, as in the case of $D$ and $B$ mesons. For heavy quarks, the vertex corrections in the DSE are insignificant and the vertex is almost bare, which is 
another reason the leading truncation is also successful for quarkonia. To see this, consider the Ball-Chiu ansatz for the quark-gluon vertex, 
\begin{equation}
 \Gamma_\mu^\mathrm{BC} (k,p) =   \tfrac{1}{2} \big [ A(k) + A(p) \big ]  \gamma_\mu  +   \frac{A(k) - A(p)}{2(k^2-p^2)} \  (\feynslash k+ \feynslash p) \, (k_\mu + p_\mu) \  
             - i \, \frac{B(k) - B(p)}{ k^2-p^2 } \ (k_\mu + p_\mu)\, \mathbb{I}_D \ ,
\end{equation}
where $A(p^2)$ and $B(p^2)$ are the vector and scalar components, respectively,  of the quark's gap equation solution: $S^{-1} (p) =  i A(p^2)\, \feynslash p   + B(p^2)\, \mathbb{I}_D$.
As the mass and wave functions, $M(p^2) = B(p^2)/A(p^2)$ and  $Z(p^2)= 1/A(p^2$), vary insignificantly for the $b$ quark and modestly for the $c$ quark, one 
has $A(k^2) \approx A(p^2)\simeq 1$ and $B(k^2) \approx  B(p^2)$,  where $k$ is the incoming and $p$ the outgoing quark momentum related by the gluon momentum $q=p-k$. 
In this heavy-mass limit, the tensor structures of the Ball-Chiu proportional  to the dressing functions $\Delta B(k,p)$ and $\Delta A(k,p)$ vanish~\cite{Ivanov:2007cw} and only the 
leading term survives:
\begin{equation}
  \Gamma_\mu (k,p) \ \propto \ \frac{A(k) + A(p)}{2}\  \gamma_\mu  \ \approx \  \gamma_\mu  \ .
\end{equation}
Hence, unlike the equal-mass case, the kernel of a heavy-light $\bar Qq$ meson contains very different corrections to either quark-gluon vertex. As a consequence, the repulsive 
crossed-box contribution and the two asymmetrically dressed vertices do not cancel perfectly. The rainbow-ladder truncation is therefore not a good approximation and one must take 
the unavoidable dressing of the light quark seriously, which adds to the angular complexity of the $\bar Qq$ scattering kernel with terms proportional to $z_k$ discarded in the lowest 
truncation. It is plausible that this additional angular dependence of the kernel will cancel imaginary contributions to Chebyshev moments by which the rainbow-ladder
truncation is plagued and thereby restore its Hermiticity.

A different case is the quark-diquark Faddeev equation employed in Refs.~\cite{Oettel:2000jj,Cloet:2008re,Segovia:2015hra}, which relies on more model input and its kernel cannot 
be described by a simple ladder approximation. More precisely, the fact that the Faddeev amplitude contains a diquark in the $\bar 3_c$ channel that is free at spacelike momenta 
but pole-free on the timelike axis implies a $qq$ scattering kernel beyond ladder truncation~\cite{Bender:1996bb}. It is however possible that the appearance of an imaginary component
in the third eigenvalue can also be traced back to missing angular dependence in the Faddeev kernel that originates in three-body interactions or additional diquark contributions
not included in Ref.~\cite{Segovia:2015hra}.

\section{Analytical quark structure: numerical DSE solutions and complex-conjugate poles \label{sec3}}

As we have argued above, while the rainbow-ladder truncation is successful in studies of meson spectroscopy and decays for a broad range of ground-state mesons, the approximation 
inexorably runs into trouble when applied to excited mesons states, heavy-light mesons and certain mass splittings, such as the $a_1$-$\rho$ mass difference~\cite{Chang:2011ei}. 
As mentioned, progress in calculation of the quark's DSE beyond the rainbow-ladder truncation has been made and is underway~\cite{Rojas:2013tza,Matevosyan:2006bk,Chang:2009zb,
Chang:2011ei,Fischer:2009jm,Qin:2013mta,Aguilar:2014lha,Binosi:2014aea,Binosi:2016rxz}. More effort is required, though, as practical computations of the Bethe-Salpeter amplitude imply
the knowledge of the quark propagator in the complex plane\footnote{See discussion in Section~III of Ref.~\cite{Rojas:2014aka} and references therein.}, where poles produced by 
a given interaction pose severe numerical difficulties. This also occurs for vertex ans\"atze beyond the rainbow-ladder ansatz. An economic expedient to represent these propagators 
by analytical expressions~\cite{Bhagwat:2002tx} is based on a complex-conjugate pole model and has been successfully used in calculations of pion distribution amplitudes and elastic form factors:
\begin{equation}
    S(p) = \sum_i^n \left [ \frac{z_i}{i\feynslash  p + m_i }  + \frac{z_i^*}{i\feynslash  p + m_i^*}\right ] \quad ,  \qquad  m_i, z_i \in \mathbb{C} \ .
    \label{pole}
\end{equation}
In essence, the expression in Eq.~(\ref{pole}) does not produce poles on the real timelike axis and has no K\" all\' en-Lehmann  representation; therefore the quark cannot appear 
in the Hilbert space of observable states. This is consistent with confinement, though in real QCD many more poles or cuts may characterize the analytic structure of the quark's 
dressing functions. The numerical DSE solutions for the quark can be fitted with $n=3$ complex-conjugate poles, where the pole locations depend on the model for the gluon- and 
vertex-dressing functions and whether the parameters $m_i$ and $z_i$ are fitted to the DSE solutions on the real axis or to a numerical solution in a parabola on the complex plane.  

We here present the results of such fits for three different cases, namely the quark propagator obtained in rainbow-ladder truncation with two different interaction parameter sets 
 listed in Table~I of Ref.~\cite{Rojas:2014aka} (corresponding to model 1 and 2 which best reproduce ground and excited states, respectively)   and with the Ball-Chiu vertex. 
As mentioned previously, the quark propagators, $S(k\pm\eta_\pm P)$,  involve complex arguments for their respective momenta in the Euclidean formulation of the BSE~(\ref{BSE}). 
It was argued that a good model to reproduce the quark propagator on the real spacelike axis and to analytically continue it on the complex plane is given by the expression in Eq.~(\ref{pole}). 
Nonetheless, using Cauchy's integral method, one can also directly solve the quark DSE on the complex plane with a given model for the gluon dressing function~\cite{Rojas:2014aka,
Fischer:2005en,Krassnigg:2009gd}. 

It is thus of interest to study whether this complex solution is in qualitative and quantitative agreement with the pole-model fit to the real-axis solution and whether the analytic structure 
of the quark-dressing functions off the real axis matters at all for hadronic observables. The latter point deals with the physical content of these poles --- if any of the fits or extrapolations 
to the complex plane yield quark propagators that produce in conjunction with the BSE exactly the same values for meson observables, may one simply view them as models that 
work well, very much like the models~\cite{Cloet:2008re,ElBennich:2010ha,Ivanov:2007cw}  based on  entire functions?

\begin{figure}[t!]
\vspace*{4mm}
\centering
\includegraphics[scale=0.42]{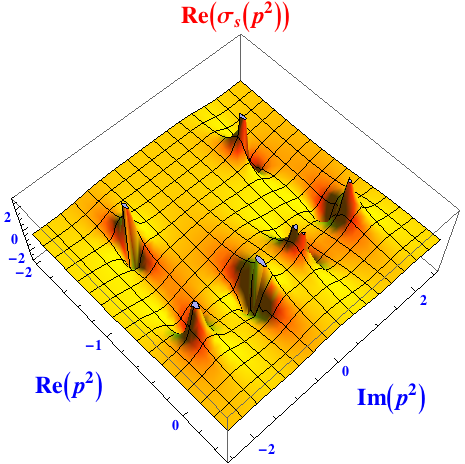} \hfill  \includegraphics[scale=0.38]{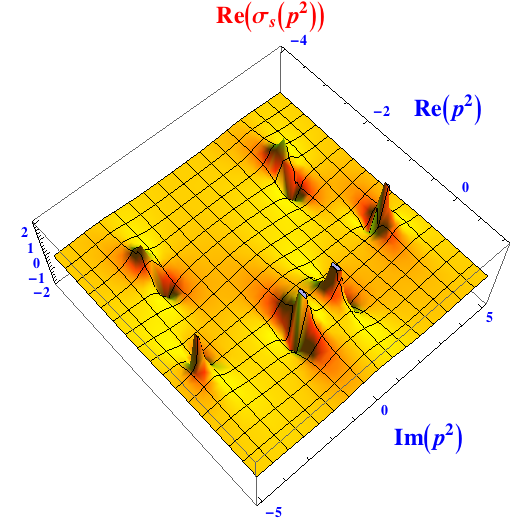} \\ \vspace*{8mm}\hspace*{-1.1cm}
\includegraphics[scale=0.47]{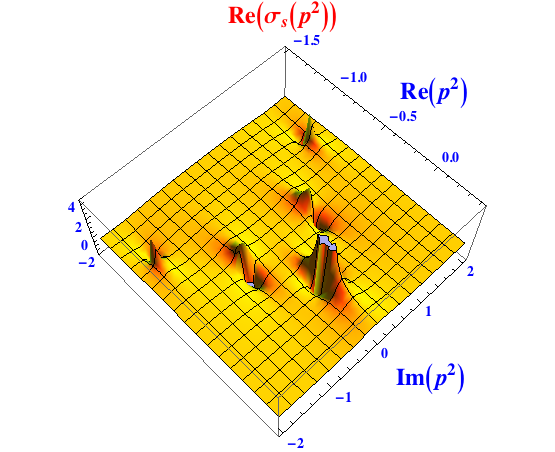} \hfill \hspace*{-1cm}\includegraphics[scale=0.44]{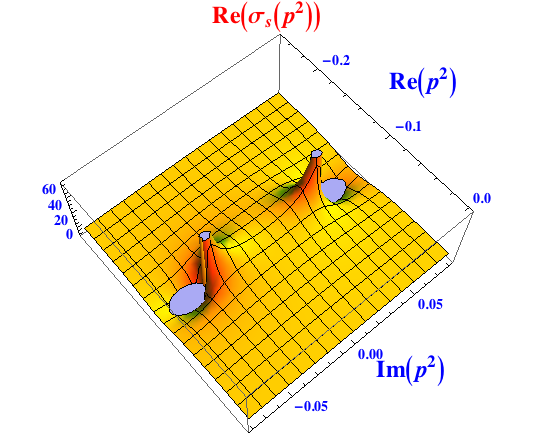} 

\vspace*{8mm}
\caption{Scalar function $\sigma_s$ of the light quark propagator $S(p) = - i \, \feynslash p \, \sigma_V (p^2)+ \sigma_ S (p^2)$ for the three cases: model~1 
in Ref.~\cite{Rojas:2014aka} with $\omega D= (0.8~\mathrm{GeV})^3$, $\omega = 0.4$~GeV, in rainbow-ladder truncation (top-left graph);  model~2 in Ref.~\cite{Rojas:2014aka} 
with $\omega D= (1.1~\mathrm{GeV})^3$,  $\omega = 0.6$~GeV, in rainbow-ladder truncation (top-right graph); Ball-Chiu vertex with  $\omega D= (0.55~\mathrm{GeV})^3$, 
$\omega = 0.5$~GeV (bottom-left and -right graph). In all cases the interaction of  Ref.~\cite{Qin:2011dd} is used with a renormalized running quark mass $m(19~\mathrm{GeV}) = 3.4$~MeV.  }
\label{fig1}
\end{figure}

The complex-conjugate pole parameters, following Eq.~(\ref{pole}) ($i=1,2,3$), for the three different interaction ans\"atze and vertices are found with a best least-squares 
fit (all entries are in GeV$^2$):
\begin{table}[h]
\begin{center}
\def\arraystretch{1.5}
\begin{tabular}{l|cc|cr|cc}
& $\operatorname{Re} m_1^2$ & $\operatorname{Im} m_1^2$ & $\operatorname{Re} m_2^2$ &  $\operatorname{Im} m_2^2$ & $\operatorname{Re} m_3^2$ & $\operatorname{Im} m_3^2$  \\
  \hline
  Model~1~\cite{Rojas:2014aka}  & $-0.24$ &  $0.53$ & $-1.35$  & $-1.37$  &  $-0.35$  & $-1.71$  \\
  Model~2~\cite{Rojas:2014aka}  & $-0.39$  & $0.83$  &  $-2.70$ &  $-2.92$  & $-0.96$  &  $-3.90$ \\
  Ball-Chiu    &  $-0.13$  &  $0.05$  &  $-0.57$  &  $0.70$  &  $-1.07$ & $-1.63$  \\ \hline
\end{tabular}
\end{center}
\label{default} 
\end{table}

\begin{figure}[t!]
\centering
 \includegraphics[scale=0.4]{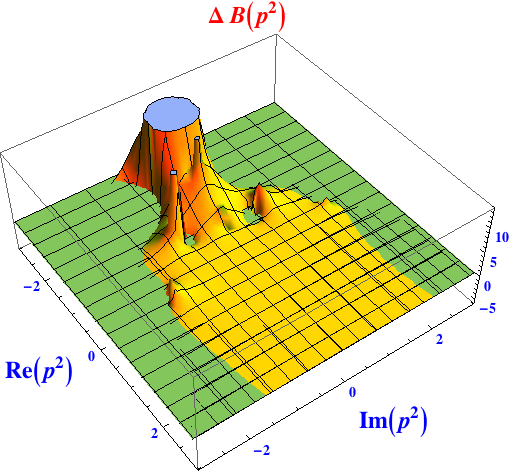}   \hfill \includegraphics[scale=0.4]{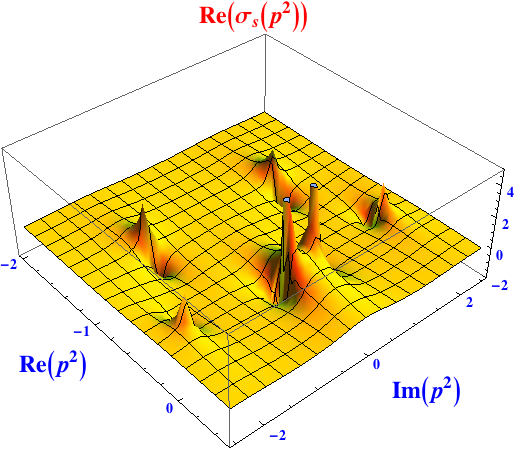}  
	\vspace*{6mm}
\caption{Left panel: Plot of the function $\Delta B(p^2) =  \operatorname{Re} B_\textrm{Model 2}(p^2) - \operatorname{Re} B_\textrm{Model 1}(p^2)$, where the green horizontal plane indicates the zero level; 
see text for details.    Right panel: complex-conjugate pole representation of $\sigma_s$ obtained with the Maris-Tandy model~\cite{Maris:1997hd}, the parameter values $\omega = 0.4$~GeV,
$\omega D = 0.372~\mathrm{GeV}^3$ and a running quark mass renormalized with $m(19~\mathrm{GeV}) = 3.4$~MeV. }
\label{fig2}
\vspace*{-1mm}
\end{figure}

\noindent
The dressed-quark propagator is the solution of a gap equation and has the general form,
\begin{equation}
S(p) = - i \,  \sigma_V (p^2)\, \feynslash p + \sigma_ S (p^2)  \mathbb{I}_D =  \left [  i A(p^2)\, \feynslash p   + B(p^2)\, \mathbb{I}_D \right ]^{-1} .
\end{equation}
To visualize the poles in the above table we plot the real component of the scalar function $\sigma_s(p^2)$ for a light quark in Fig.~\ref{fig1}. The upper row of graphs corresponds to 
the interaction parametrization~\cite{Qin:2011dd} in rainbow-ladder approximation that best reproduces the ground-state spectrum with $\omega D= (0.8~\mathrm{GeV})^3$, 
$\omega = 0.4$~GeV, and the one that yields the most satisfying excited-state spectrum with $\omega D= (1.1~\mathrm{GeV})^3$,  $\omega = 0.6$~GeV. Here, $\omega$ plays the 
role of the interaction width while $D$ accounts for its strength. Both parameters are chosen for a fixed value of $\omega D$ and describe effectively the combined effect of 
the gluon and vertex dressing functions in the leading truncation and therefore determine the support and location of maximal strength of the interaction. Comparing both graphs, 
one observes a set of two conjugate-complex poles that are located in the vicinity of the real axis. The two other pairs lie more remote from the real axis, but in the excited-state 
parametrization this distance increases considerably and the poles shift towards much larger timelike squared momenta. When the gap equation is solved with the  
beyond rainbow-ladder Ball-Chiu ansatz for the vertex, one of the complex-conjugate pole pairs moves indistinguishably close to  the real axis, whereas the pole pair most 
distant from the  real axis shifts toward larger timelike squared momenta as depicted in the bottom-left graph of Fig.~\ref{fig1}. In the bottom-right graph, we plot the enlarged 
vicinity  of the complex-conjugate poles closest to the real axis for better visibility. 
The left panel of Fig.~\ref{fig2} depicts the difference of the real part of the excited state's mass function, $\operatorname{Re} B_\textrm{Model 2}(p^2)$, and the ground
state's mass function, $\operatorname{Re} B_\textrm{Model 1} (p^2)$: $\Delta B(p^2) = \operatorname{Re} B_\textrm{Model 2}(p^2) - \operatorname{Re} B_\textrm{Model 1}(p^2)$.
Upon inspection of the real function $\Delta B(p^2)$ in the complex plane it becomes clear that the most dramatic differences are localized in the timelike region close to the complex-conjugate
poles which characterize the different parameterizations of the given interaction~\cite{Qin:2011dd}. Below the zero level, represented by a green-colored plane, $\Delta B(p^2)$ 
becomes slightly negative but especially in the spacelike region the difference is vanishing. For further comparison, we reproduce in the right panel of Fig.~\ref{fig2} again the  function 
$\sigma_s(p^2)$  obtained with the Maris-Tandy interaction~\cite{Maris:1997hd} from which it can be inferred that in a fit with three conjugate-complex poles one pole pair lies again closer
to the real axis. Roughly, this solution gives rise to a pole structure that resembles the ground-state parametrization in the top-left corner of Fig.~\ref{fig1}.

\section{Final remarks}

At the present point it is difficult to draw firm conclusions about the analytic structure of the quark propagators we just presented. It is evident that either complex-conjugate pole 
representation is too simplistic: we checked that while they yield the same numerical results for light-meson observables as the full DSE solution in a parabola on the complex 
plane~\cite{Rojas:2014aka,Fischer:2005en,Krassnigg:2009gd}, they do so only for a restrained set of hadronic data. It is reasonable to assume that a realistic solution, such as 
the one produced by fully unquenched QCD, is considerably more complicated. We note  that the pole model also works well for charm quarks and we verify that the $D$ and $D_s$ 
masses and decay constants are almost indistinguishable from the ones obtained with the numerical solution in a complex parabola. The effect of all 12 tensor structures derived from 
longitudinal {\em and\/} transverse Slavnov-Taylor identities (STI)~\cite{He:2009sj}  in the quark-gap equation is currently being studied~\cite{STIProject}, which calls for a full 
comparison of the analytical structure of the quark propagators in the rainbow-ladder approximation with that including all vertex tensor structures.

\begin{acknowledgements}
 B.~E. thanks the organizers of the  ECT* workshop ``Nucleon Resonances From Photoproduction to High Photon Virtualities"  for their kind invitation and local support. 
 The work mentioned in this contribution was made possible by: FAPESP grant nos.~2013/01907-0 and 2013/16088-4 (S\~ao Paulo State); CNPq grant nos.~305894/2009-9,
 458371/2014-9 and 305852/2014-0 (Brazil). 
E.~R. acknowledges support by {\em Patrimonio Aut\'onomo Fondo Nacional de Financiamiento para la Ciencia, la Tecnolog\'ia y la Innovaci\'on, Francisco Jos\'e de Caldas\/} 
and by {\em Sostenibilidad-UDEA  2014--2015\/} (Colombia). 
\end{acknowledgements}

\end{document}